\documentclass[showpacs,twocolumn,amsmath,amssymb,fixfloats]{revtex4}
\usepackage[english]{babel}
\usepackage{amsmath}
\usepackage{amssymb}
\usepackage{graphicx}

\newcommand{\ket}[1]{|#1\rangle}

\begin{document}

\title{Cavity QED characterization of many-body atomic states in double-well potentials -- The role of dissipation}

\author{W. Chen}
\author{P. Meystre}

\affiliation{B2 Institute, Department of Physics and College of
Optical Sciences,\\
The University of Arizona, Tucson, AZ 85721 }

\begin{abstract}
When an incident light beam is scattered off a sample of ultracold
atoms trapped in an optical lattice, the statistical properties of
the retro-reflected field contain information about the quantum
state of the atoms, and permit for example to distinguish between
atoms in a superfluid and a Mott insulator state. This paper
extends our previous analysis of this problem to include the
effects of cavity damping. We use a Monte Carlo wave function
method to determine the two-time correlation function and
time-dependent physical spectrum of the retro-reflected field in
the case of a simple two-well lattice. We also analyze
quantitatively the entanglement between the atoms and the light
field for atoms in a Mott insulator and a superfluid state.

\end{abstract}
\maketitle

\section{\label{introduction} Introduction}

Ultracold atoms in optical lattices provide a remarkable test system to simulate a number
of situations in condensed matter physics \cite{Lewenstein:MimickingCondensedMatter}
 under exquisitely controlled conditions. The first example along these lines was the
superfluid to Mott insulator transition \cite{Greiner:QuantumPhaseTransitionSF2MI, Greiner:CollapseRevivalofMatterWaveFieldofBEC}
 in bosonic systems, but the list of strongly correlated condensed matter systems
that can be simulated by atoms or molecules
\cite{Rom:MoleculesLattices,Stoeferle:MoleculesLattice,Thalhammer:MoleculesLattice,Jaksch:creationmoleculeinOL}
in optical lattices has continued to grow since these pioneering experiments. Examples include,
but are not limited to the Bose-Hubbard and Fermi-Hubbard models \cite{Jaksch:ColdBosonicAtomsinLattice, Zoller:OLreview},
spin systems \cite{GarciaRipoll:SpinHamiltonians} and the Anderson
lattice model \cite{Miyakawa:AndersonModel}. In other examples, rotating lattices
\cite{Barberan:OrderedStructuresInRotatingUltracoldBoseGas} are expected to lead to the
realization of analogs of the quantum Hall effect, and random lattices have recently been
used to study Anderson localization in atomic systems \cite{Roati:AndersonLocalizationBEC}.

The ability to characterize the many-particle state of the atomic fields in optical lattices 
is of course central to the realization of these experiments. Some schemes are designed for probing
those states nondemolitionly 
\cite{Mekhov:LightScatteringFromColdAtoms, Eckert:QuantumNondemolitionDetectionOfStronglyCorrelatedSystems}.
In a recent paper \cite{Chen:CavityQEDdetermination}, we proposed an optical scheme based on
the diffraction of a quantized light field off the atomic sample to probe the number statistics 
of the matter-wave field. The basic idea is to use two light fields counterpropagating 
in a high-$Q$ ring cavity and coupled via Bragg scattering off the atoms, a technique 
that is analogous to the Bragg reflection of X-rays off a crystal, but operating in the 
quantum regime. We found that the dynamics of the light field strongly depends on the 
manybody state of the atomic field as well as on the lattice spacing. Specifically, 
the statistical properties of the Bragg-reflected light field for atoms in a Mott insulator 
state and for a superfluid described both in terms of a number-conserving state and a 
mean-field coherent state were found to provide a clear signature of the state of the atomic field.

The present paper extends these results to include the effects of cavity damping.
We use a Monte Carlo wave function method \cite{Molmer:MCWFmethodinQO} to determine the
two-time correlation functions and (time-dependent) physical spectrum \cite{Eberly:PhysicalSpectrum}
of the retro-reflected field in the case of a simple two-well lattice. We also analyze
quantitatively the entanglement \cite{Vidal:ComputableMeasureEntanglement}
between the atoms and the light field for atoms in a Mott insulator and a superfluid state.
Our main result is that even in the presence of dissipation these quantities allow to easily
distinguish between a Mott insulator and a superfluid atomic state, and in addition permit
to decide between two familiar descriptions of that state.

Section II describes the main elements of our model, introducing in particular the effective 
non-hermitian Hamiltonian required in the Monte Carlo wave function simulations of the problem. 
The intensity reflected by the atoms via Bragg scattering is discussed in section III, 
and section IV presents the time-dependent physical spectrum of the reflected light. 
The quantum entanglement that may develop as a result of Bragg diffraction is quantified 
in section V in terms of the logarithmic negativity. Finally, section VI is a conclusion and outlook.

\section{\label{model} Model}

\begin{figure}
\includegraphics[width=60mm]{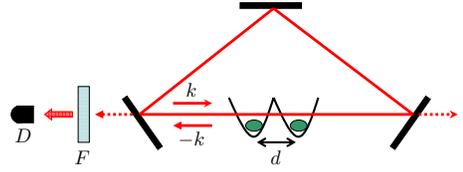}
\caption{Atoms trapped in an optical lattice with lattice constant
$d$ and interacting with two counter-propagating
modes with wave vectors $\pm k$ in a ring resonator. F: filter. D: Detector.}
\label{schematic}
\end{figure}

We consider a sample of ultracold bosonic two-level atoms \cite{Jaksch:ColdBosonicAtomsinLattice}
with transition frequency $\omega_a$ trapped in the lowest Bloch band of a one-dimensional double-well potential
\cite{Salasnich:BECinDoubleWellTrap, Thomas:DoubleWellMagneticTrapforBEC, Milburn:QuantumDynamicsofBECinDoubleWell}
placed inside an optical ring resonator. The atoms are driven by two counterpropagating
cavity modes of wave vectors $\pm k$ and
frequency $\omega_k= kc$, see Fig.~\ref{schematic}. This interaction is described by the dipole
interaction Hamiltonian
\begin{equation}
    \mathcal{\hat V}_d= \int dx \hat \psi_e^\dagger(x)
    \langle e| {\hat {\bf d}}\cdot {\hat {\bf E}}(x)|g
    \rangle \hat \psi_g(x)+ h.c.
\end{equation}
where $\hat {\bf d}$ is the dipole moment of the transition,
$\hat {\bf E}(x)$ is the electric field operator, and $\hat \psi_e$ and $\hat \psi_g$ 
are field operators describing atoms in their excited and ground
electronic states $|e\rangle$ and $|g\rangle$, respectively. We
assume that the optical fields are sufficiently detuned from the
atomic transition frequency $\omega_a$ that the excited electronic
state can be adiabatically eliminated.

We proceed by expanding $\hat \psi_e$ and $\hat \psi_g$ on the Wannier basis
of the lowest Bloch band as
\begin{equation}
    \hat \psi_{e,g}(x)=\sum_{m}
    \psi_m^{(e,g)}(x)\hat c_m^{(e,g)},
\end{equation}
where $\hat c_m^{(e)}$ and $\hat c_m^{(g)}$ are the annihilation
operators for excited and ground state atoms in the
$m^{\rm th}$ well, and $\psi_m^{(e)}$ and $\psi_m^{(g)}$ are the
corresponding wave functions. Introducing also the operator
\begin{equation}
\hat N(d)=\hat n_0 + \hat n_1 e^{2ikd},
\label{nd}
\end{equation}
where
\begin{equation}
\hat n_m \equiv \hat c_m^{(g)\dagger} \hat c_m^{(g)},
\end{equation}
and $d$ is the well separation, the atom-field system is
easily seen to be described by the Hamiltonian \cite{Chen:CavityQEDdetermination}
\begin{eqnarray}
\hat H&=& \sum_k \hbar \omega_k \hat a_k^\dagger \hat a_k+ \hbar
g[\hat N(0)(\hat a_k^\dagger \hat a_k
+\hat a_{-k}^\dagger \hat a_{-k}) \nonumber \\
 &+& \hat N(d) \hat a_{-k}^\dagger\hat a_k
 +\hat N(-d)\hat a_k^\dagger \hat a_{-k}]
\label{effectiveh1}
\end{eqnarray}
where
$$
    g=\frac{2\wp^2}{\Delta \hbar^2}\left|\int dx \mathcal
    E_{\pm k}(x)\psi_{0}^{(e)*}(x)\psi_0^{(g)}(x)\right|^2,
$$
and $\wp$ is the dipole matrix element of the atomic transition. We drop the label $(g)$ 
in the following for notational clarity since no confusion is possible once the excited electronic state has been eliminated.

Cavity damping is treated in the usual way by coupling the cavity
modes to a Markovian reservoir of harmonic oscillators of
frequencies $\{\omega_q\}$, with the interaction Hamiltonian
\begin{equation}
\mathcal{\hat V}_r =\hbar (\hat a_k + \hat a_{-k})\sum_q g_q \hat
b_q^\dagger + h.c.
\end{equation}
Finally,  the $+k$ cavity mode is driven by an oscillating
classical current of amplitude $\eta$ and frequency
$\omega=\omega_k$,
\begin{equation}
\mathcal{\hat V}_p=\hbar \eta e^{-i \omega_k t} \hat a_k^\dagger+
h.c.
\end{equation}

The numerical studies that are the subject of this paper are
conveniently carried out using a Monte Carlo wave function approach
\cite{Molmer:MCWFmethodinQO}.
In this case, the coupling of the system to the reservoir is
described in terms of the effective non-hermitian Hamiltonian
\begin{equation}
\hat H_{\rm eff}=\mathcal{\hat H}_S - \frac{i \hbar}{2}
\sum_{j=1,2} \hat{C}_j^\dagger \hat{C}_j,
\end{equation}
where
\begin{equation}
\hat{C}_1=\sqrt{2 \gamma} \hspace{0.1cm}\hat a_k, \hspace{0.5cm}
\hat{C}_2=\sqrt{2 \gamma} \hspace{0.1cm} \hat a_{-k}
\end{equation}
and $\gamma$ is the familiar Wigner-Weisskopf decay rate resulting
from the coupling of the cavity modes to the Markovian reservoir.
For the specific situation at hand this gives, in an interaction
picture with respect to the Hamiltonian ${\hat H}_0= \sum_k\hbar
[\omega_k + g\hat N(0)]\hat a_k^\dagger \hat a_k$,
\begin{eqnarray}
\hat H_{\rm eff} &=& \hbar \left [g \hat N(d) a_{-k}^\dagger a_k
+ \eta \hat a_k^\dagger e^{ig \hat N(0) t} + h.c. \right ]\nonumber \\
 &-& i \hbar \gamma \left [\hat a_k^\dagger \hat a_k + \hat a_{-k}^\dagger \hat
 a_{-k} \right ].
\label{effectiveh}
\end{eqnarray}

\section{\label{Reflected intensity} Reflected intensity}

The light intensity transmitted and reflected by the trapped atoms
depends strongly on the state of the atoms as well as on the well
separation $d$. Throughout this paper we assume that the
incident $+k$ mode is initially in a coherent state (with amplitude
$\alpha=\sqrt{2}$ in our simulations) and the reflected $-k$ mode is in the vacuum state, 
and we concentrate on the special cases of well separations $d=\lambda/4$ and $\lambda/2$, 
for which the Hamiltonian $\hat H_{\rm eff}$ becomes
\begin{eqnarray}
\hat H_{\rm eff}^{(\pm)} &=& \hbar g(\hat n_0\pm \hat n_1)
\left ( a_{-k}^\dagger a_k + a_k^\dagger a_{-k} \right )  \nonumber \\
&+& \hbar \eta \left (\hat a_k^\dagger e^{i g(\hat n_0+\hat n_1)t} + h.c. \right ) \nonumber \\
&-& i \hbar \gamma \left ( a_k^\dagger a_k + a_{-k}^\dagger a_{-k} \right ).
\label{effectivehpm}
\end{eqnarray}
Here the `plus' sign corresponds to the $d=\lambda/2$ case and the
`minus' sign to the $d=\lambda/4$ case.

\subsection{Mott insulator}

When in a Mott insulator state the atomic field has a well-defined
number of atoms in each well, and thus can be described by a product of Fock states. 
Atomic Fock states are eigenstates of the effective Hamiltonian
(\ref{effectivehpm}), that is, the operators $\hat n_0$ and $\hat n_1$
are constants of motion with eigenvalues $n_0$ and $n_1$, and it
is possible to replace the operators by these eigenvalues in
Eq.~(\ref{effectivehpm}).

It is known \cite{Glauber:ClassicalBehaviorSystemsQO}
that if a system of coupled harmonic oscillators satisfies Heisenberg
equations of motion that can be expressed as
\begin{equation}
\dot{\hat{a}}_j=F_j(\{\hat a_k(t)\}, t), \,\,\,\,\, j=1 \ldots n,
\end{equation}
where the functions $F_j$ may depend explicitly on time, then if
the oscillators are initially in a coherent state they will remain
in a coherent state for all times. This is the case for the
situation at hand. Hence the two modes of the light field are in
coherent states whose amplitudes exhibit damped oscillations
due to the combined effects of photon exchange between the incident and reflecting 
modes and of cavity decay. As is to be expected, the oscillation frequency is proportional 
to the total number of atoms for the $d=\lambda/2$
case and to the difference in the populations of the two wells for
$d=\lambda/4$. The difference between the oscillation frequencies in these
two cases permits therefore a full determination of the well
populations.

\begin{figure}
\includegraphics{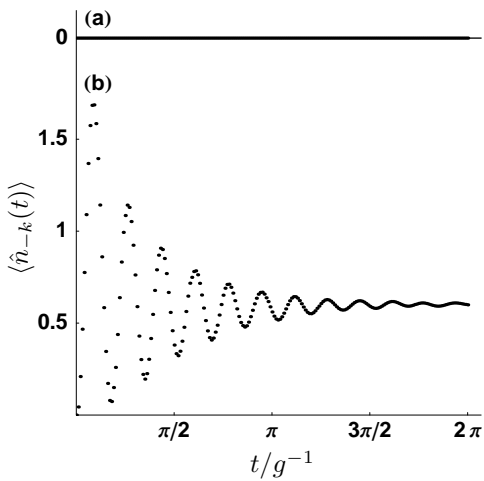}
\caption{Single trajectory (or average reflected intensities
$\langle \hat n_{-k}(t)\rangle$) for atoms in an Mott insulator state
with 3 atoms per well for (a) $d=\lambda/4$ and (b)
$d=\lambda/2$, $\eta=1.5g$ and $\gamma=0.9g$. }
\label{nmkfock}
\end{figure}

Figure \ref{nmkfock} illustrates  the time dependence
of the reflected intensity for equal well populations  and for the two
special well separations. Different atomic populations in the two wells also result in an oscillating signal for the
``destructive well separation'' $d=\lambda/4$, but they are is much weaker
and slower than that in the ``constructive case'' $d=\lambda/2$.

We mentioned that the light field remains in a
coherent state at all times. These are eigenstates of the
annihilation operator, hence quantum jumps do not affect them, and
all Monte Carlo Wave function trajectories are identical in this
case.

\subsection{Superfluid}
There are two particularly simple ways to approximate the state of the atoms 
in a superfluid state. The first one is essentially a mean-field approach that assumes
that the atoms in the two wells are in coherent states,
\begin{equation}
\ket{\psi_{\rm SF1}}=\ket{\alpha_0,\alpha_1}=\sum_{n_0,n_1} c_{n_0,n_1} \ket{n_0,n_1},
\label{sf1}
\end{equation}
whereas the second description accounts for the fixed total
number $N$ of atoms and describes their state as
\begin{eqnarray}
\ket{\psi_{\rm SF2}}&=&\mathcal N^{-1} (c_1^{\dagger}+c_2^{\dagger})^N\ket{0,0}  \nonumber \\
&=&\sum_{n_0}^N  \sqrt{\frac{N!}{2^N n_0! (N-n_0)!}}\ket{n_0, N-n_0}  \nonumber \\
&=&\sum_{n_0}^N  b_{n_0} \ket{n_0, N-n_0}.
\label{sf2}
\end{eqnarray}
This section compares the reflected optical field corresponding to these two
descriptions, assuming as before that the incident light field is initially in a 
coherent state and the reflected field is in a vacuum state.

\begin{figure}[t]
\includegraphics{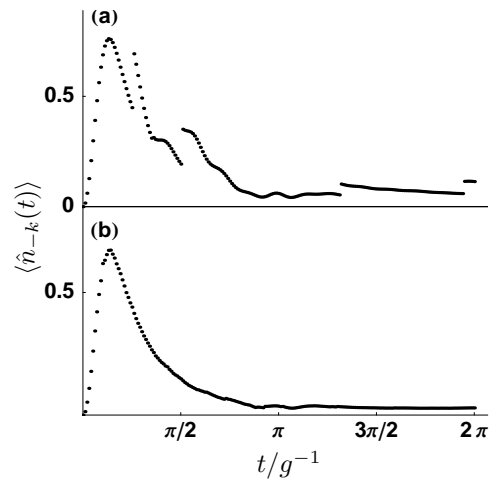}
\caption{Reflected intensity $\langle \hat n_{-k}(t)\rangle$ for
atoms initially in the superfluid state $\ket{\psi_{\rm SF1}}$
with 4 atoms on average per well and $d=\lambda/4$.
(a) typical single trajectory. (b) average over 100 trajectories, for $\eta=1.5g$ and $\gamma=0.9g$.}
\label{nmkco}
\end{figure}

Consider first the coherent state description of Eq.~(\ref{sf1}).
From the discussion of the Mott insulator case, we know that for
fixed atom numbers in the two wells the state of the field remains
a coherent state,
$$
\ket{\alpha,0;n_0,n_1} \rightarrow \ket{\alpha_k(n_0,n_1,d,
t),\alpha_{-k}(n_0,n_1,d,t); n_0,n_1},
$$
so that without quantum jumps between times 0 and $t$
\begin{eqnarray}
&&\ket{\psi(t)} = \nonumber \\
&&\sum_{n_0,n_1} c_{n_0,n_1}
\ket{\alpha_k(n_0,n_1,d,t),\alpha_{-k}(n_0,n_1,d,t);n_0,n_1}.\nonumber \\
\label{copsi}
\end{eqnarray}
It is apparent from this expression that the atoms and the light
field become entangled, a point to which we return in section V.
Also, since the optical field does not generally remain in a coherent state
in this case, the quantum jumps resulting from the loss of photons
due to dissipation become apparent in the single trajectories, see
Fig.~\ref{nmkco}.

\begin{figure}[t]
\includegraphics{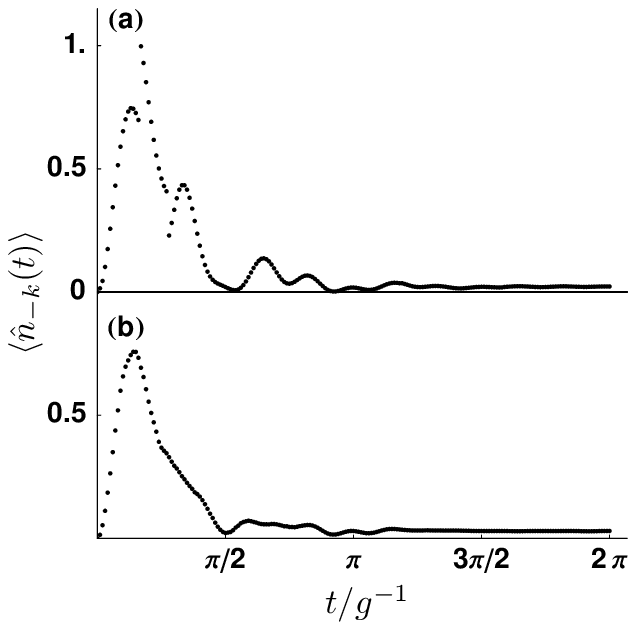}
\caption{Reflected intensity $\langle \hat n_{-k}(t)\rangle$ for
atoms initially in the superfluid state $\ket{\psi_{\rm SF2}}$
with a total of 8 atoms and $d=\lambda/4$.
(a) typical single trajectory; (b) average over 100 trajectories, for $\eta=1.5g$ and $\gamma=0.9g$.}
\label{nmksf2}
\end{figure}

Comparing the present case to the situation for a Mott insulator
state, we observe that there are no contributions to the
backscattered intensity from situations with equal populations in
the two wells, since the coupling between the incident and
reflected fields is zero in these cases. Also, the oscillatory
character of the back-reflected light is now largely washed out by
the uncertainty of the atom numbers in the two wells.

The number-conserving description (\ref{sf2}) of the superfluid
state leads to similar results, although the details of the
dynamics are slightly different because of the fixed total number of
atoms which results in less atom number uncertainty, see Fig.~\ref{nmksf2}.
In practice, it is therefore not expected that the backscattered
intensity will allow to unambiguously distinguish between the two
descriptions.

\section{\label{ps} Physical spectrum}
We now turn to the analysis of the physical spectrum of the
reflected optical field, which also provides a signature of the state of
the atomic field.

Because the process under study is not stationary, the usual spectrum obtained from the 
Wiener-Khintchine theorem is not appropriate, and we use instead the time-dependent physical
spectrum of Ref.~\cite{Eberly:PhysicalSpectrum}. We recall that this spectrum is defined as
\begin{eqnarray}
S(t,\omega;\Gamma)&=&\int_{0}^{t} \int_{0}^{t}
H^*(t-t_1,\omega,\Gamma) H(t-t_2,\omega,\Gamma)  \nonumber \\
&\times&\langle \hat a_{-k}^\dagger(t_1) \hat a_{-k}(t_2) \rangle
dt_1 dt_2 ,\label{gg}
\end{eqnarray}
where $H(t,\omega,\Gamma)$ is the response function of the filter,
$\omega$ its setting frequency, and $\Gamma$ its bandwidth, see Fig.~\ref{schematic}. 
Following Ref.~\cite{Eberly:PhysicalSpectrum}, we assume that it is constant for 
the frequency range of interest, and choose the filter response function as
\begin{equation}
H(t,\omega;\Gamma)=\Theta(t) \Gamma e^{-(\Gamma +i \omega) t}
\label{response}
\end{equation}
where $\Theta(t)$ is the unit step function.

\begin{figure}[t]
\includegraphics{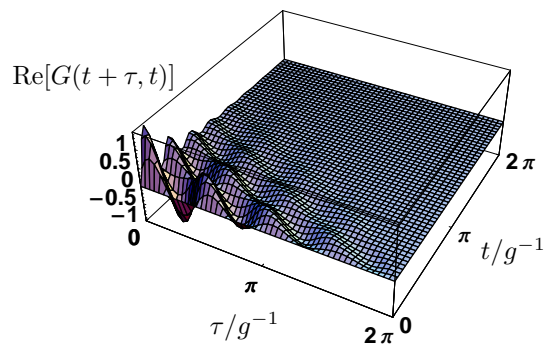}
\caption{Two-time correlation function $G(t+\tau, t)$ in the Mott
insulator regime, with 6 atoms in well 0 and 2 in well 1. Well separation is
$d=\lambda/4$, the pump constant is $\eta=0.1g$ and the decay constant is $\gamma=0.5g$.}
\label{corrfock}
\end{figure}

\begin{figure}[t]
\includegraphics{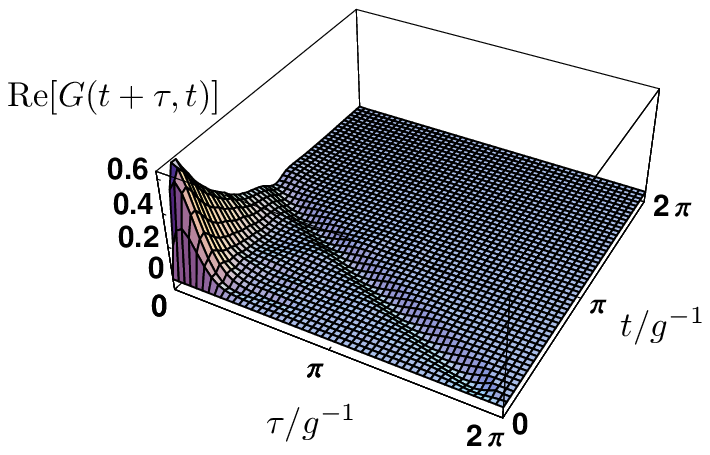}
\caption{Two-time correlation function $G(t+\tau, t)$ for atoms initially in the superfluid 
state $|\psi_{\rm SF1}\rangle$ with a mean atom number of 4 per well and  
$d=\lambda/4$, $\eta = 0.1 g$ and $\gamma = 0.5 g.$ The curve is the average over 50 
trajectories at time $t$ and 3 trajectories at time $\tau$. }
\label{corrcohe}
\end{figure}

\begin{figure}[t]
\includegraphics{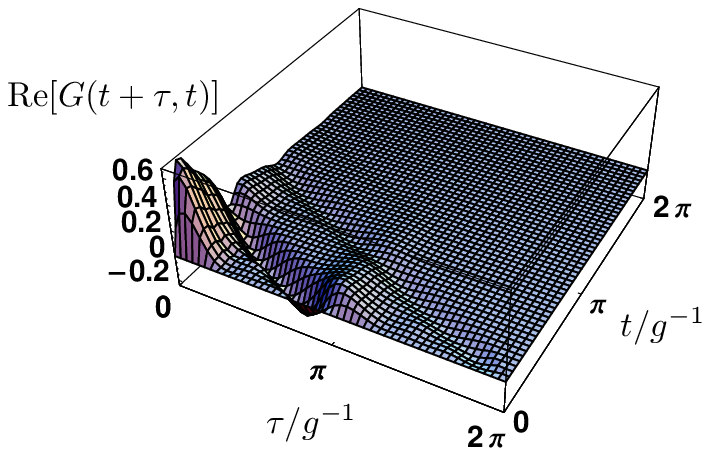}
\caption{Two-time correlation function $G(t+\tau, t)$ for atoms initially in the 
superfluid state $|\psi_{\rm SF2}\rangle$ with a total number of
8 atoms, $d=\lambda/4$ $\eta=0.1g$ and $\gamma=0.5g$. The curve is an
over 50 trajectories at time $t$ and 3 trajectories at time $\tau$.}
\label{corrsf2}
\end{figure}

Figure~\ref{corrfock} shows the real part of the two-time correlation function \begin{equation*}
G(t, t+\tau)=\langle \hat a_{-k}^\dagger(t) \hat a_{-k}(t+\tau) \rangle.
\end{equation*}
for $d=\lambda/4$ and in the range $0 \le t+\tau \le T=2\pi/g$ when the atoms are in 
a Mott insulator state, while Fig.~\ref{corrcohe} shows that same function for the initial 
superfluid state $|\psi_{\rm SF1}\rangle$, and Fig.~\ref{corrsf2} for the initial 
superfluid state $|\psi_{\rm SF2}\rangle$.  As was the case for the reflected intensity, 
the periodic oscillations of the reflected intensity characteristic of the Mott insulator 
case are washed out in the superfluid regime. Since the number-conserving description
$|\psi_{\rm SF2}\rangle$ of the superfluid state corresponds to a smaller atom number 
uncertainty, the corresponding correlation function is characterized by stronger oscillations, 
intermediate between its description in terms of coherent states and the Mott insulator case, see Fig.~\ref{corrsf2} .

\begin{figure}[t]
\includegraphics{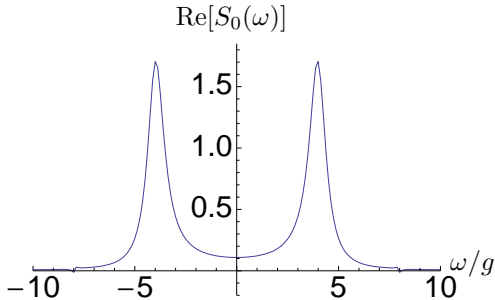}
\caption{Real part of $S_{0}(\omega)$ for a
Mott insulator states with 6 atoms in well 0 and 2 atoms in well 1
, with $d=\lambda/4$. The peak is centered at $|n_0-n_1|$ in this case.}
\label{psfock}
\end{figure}

In terms of $\tau = t_1-t_2$, the physical spectrum $S(t, \omega; \Gamma)$ may be reexpressed as
\begin{eqnarray}
S(t,\omega,\Gamma)&=&2\Gamma^2 \, e^{-2 \Gamma t} \int_{0}^{t} d\tau
 e^{(\Gamma-i\omega)\tau} \nonumber \\
&\times&\int_{0}^{t-\tau} dt_2 \, e^{2 \Gamma t_2} \, \,
{\rm Re}[\langle \hat a_{-k}^\dagger(t_2+\tau) \hat a_{-k}(t_2)\rangle] \nonumber \\
&\equiv&2\Gamma^2 \, e^{-2 \Gamma t} S_0(\omega,\Gamma).
\end{eqnarray}
with $\tau \ge 0$.

\begin{figure}
\includegraphics{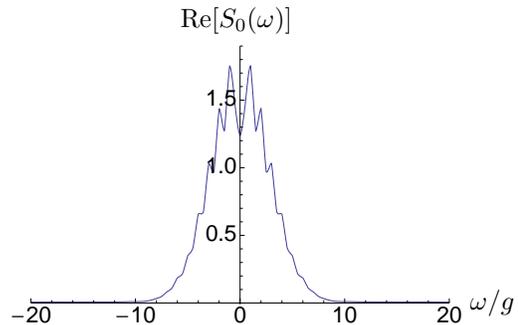}
\caption{Real part of $S_{0}(\omega)$ of
superfluid states $|\psi_{\rm SF1}\rangle$ with each well containing 4 atoms on average and $d=\lambda/4$. }
\label{pscohe}
\end{figure}

\begin{figure}
\includegraphics{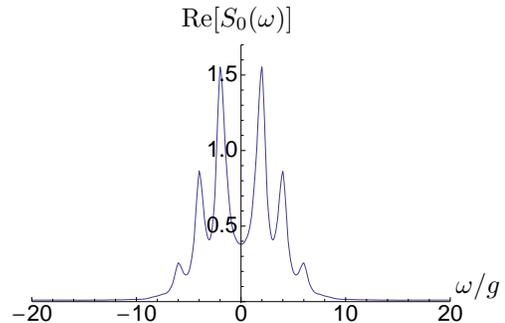}
\caption{Real part of $S_{0}(\omega)$ for the
superfluid state $|\psi_{\rm SF2}\rangle$ with a total number of 8 atoms
and $d=\lambda/4$. }
\label{pssf2}
\end{figure}

Figure~\ref{psfock} shows the real part of $S_0(\omega)$ for atoms initially in a Mott 
insulator state, while Figs.~\ref{pscohe} and \ref{pssf2} are for atoms initially in superfluid states.
In Fig.~\ref{psfock}, the temporal oscillations of the correlation function
$G(t+\tau, t)$ translates into a modulation of the lorentzian spectrum
due to the filter function at the oscillation frequency $\omega_{-} \equiv g |n_0-n_1|$ 
between the incident and reflected light fields. The physical spectrum provides therefore 
a direct measure of the population difference
between the two wells for $d= \lambda/4$, of their sum for
$d=\lambda/2$ and hence of $n_0$ and $n_1$ separately, i.e.,
\begin{eqnarray}
&&\omega_{\pm}= g(n_0 \pm n_1)  \nonumber \\
&&n_{{0,1}}= (\omega_{+} \pm \omega_{-})/2g
\end{eqnarray}
where we assumed $n_0>n_1$ for concreteness. The second small dip
at frequency $g (n_0+n_1)$ originates from the pump term,
$\hbar \eta (\hat a_k^\dagger e^{i (\hat n_0+\hat n_1)t} + h.c. )$,
in the Hamiltonian~(\ref{effectivehpm}).

Figures~\ref{pscohe} and \ref{pssf2} show the physical spectrum for the coherent and 
number-conserving superfluid states $|\psi_{\rm SF1}\rangle$ and $|\psi_{\rm SF2}\rangle$, 
respectively. In the number-conserving description (\ref{sf2}) the total atom number $N$ 
is fixed, hence for $d=\lambda/4$ the population difference between the two wells $|2 n_0-N|$ 
can only take all even or all odd numbers, depending on $N$. This restriction disappears
in the coherent state description of  $|\psi_{\rm SF1}\rangle$, so that additional peaks 
appear although the mean number of atoms in each well are the same in Figs.~\ref{pscohe} and \ref{pssf2} .

We conclude this discussion by mentioning that for a well separation $d= \lambda/2$ case 
the superfluid descriptions $|\psi_{\rm SF1}\rangle$ and $|\psi_{\rm SF2}\rangle$ result 
in completely different spectra. For $|\psi_{\rm SF1}\rangle$ the spectrum is similar to 
that of the $d= \lambda/4$ case since $n_0$ and $n_1$ are independent and hence give all 
combinations of different total numbers $(n_0+n_1)$ to form the spectrum. By contrast, 
the number-conserving description $|\psi_{\rm SF2}\rangle$ results in a single sharp peak, 
just as in the  Mott insulator case, see Fig.~\ref{psfock}.

\begin{figure}
\includegraphics{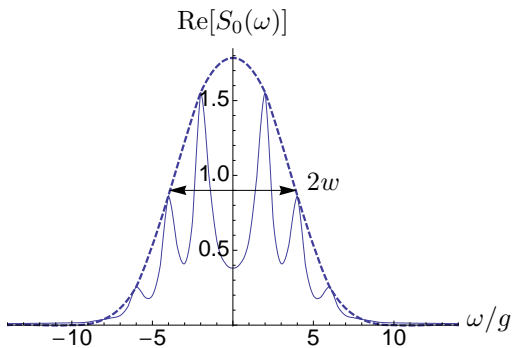}
\caption{The envelope and FWHM. }
\label{enve}
\end{figure}

\begin{figure}
\includegraphics{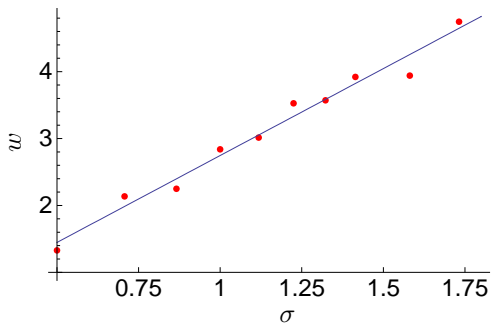}
\caption{FWHM $w$ vs. atomic number uncertainty of each well $\sigma$.
The data points corresponds to $N$=1,2,3,4,5,6,7,8,10 and 12. The straight line
is the fit to the data points. }
\label{sd}
\end{figure}

Next we use the number-conserving approximation $|\psi_{\rm SF2}\rangle$ to investigate 
the relation between the full width at half maximum(FWHM) of the physical spectrum and 
the atomic number uncertainties in the two wells. We evaluate that width with the help 
of an envelope function fitted as illustrated in Fig.~\ref{enve}. The atomic number 
uncertainties $\sigma$ are defined in the usual way as
\begin{equation}
\sigma = \sqrt{\sum_{m=0}^N (m-\bar{m})^2 ~ b_m^2} ~ ,
\end{equation}
where $\bar{m}=\sum_m mb_m^2$, $N$ is the total number of atoms, and $b_m^2$, the probability 
of having $m$ atoms in one of the wells, is a binomial distribution. Figure~\ref{sd}, 
which plots the FWHM $w$ of the physical spectrum as a function of $\sigma$, shows as expected 
a linear dependence, further illustrating the use of that spectrum in helping characterize 
the many-body state of the atomic system.

\section{\label{ent} Entanglement}
In this section we make some brief remarks on the entanglement generated between the light 
fields and the atoms, and its dependence on the state on the atomic system. We characterize 
the entanglement in terms of the logarithmic negativity \cite{Vidal:ComputableMeasureEntanglement}
\begin{eqnarray}
E_{\mathcal{N}}(\rho)=\log_{2}||\rho^{T_A}||_1,
\label{ln}
\end{eqnarray}
where $\rho^{T_A}$ is the partial transpose of $\rho$ and $||\centerdot||_1$
denotes the trace norm.

There is always some ambiguity in the way a quantum system is described in terms of subsystems. 
One simple way to describe the system at hand is as a four-partite system comprised of two optical 
modes and the atoms in the two wells. It is then possible to trace over any two parts of 
the full system, and to consider the remaining subsystems only. We find that independently 
of our choice of the subsystems being considered the resulting reduced density operator does 
not retain any trace of whatever entanglement may have characterized the full system. 
This leads us to describe instead the full system as a bipartite system, the two subsystems 
being the optical field and the atoms.

The situation is particularly simple if the atomic system is initially in a Mott insulator 
state. As discussed in Section \ref{Reflected intensity}, in the absence of quantum jumps 
between the times 0 and $t$ we have that
\begin{eqnarray}
\ket{\psi(t)}=\ket{\alpha_k(n_0,n_1,d, t), \alpha_{-k}(n_0,n_1,d, t); n_0,n_1}
\label{psitfock}
\end{eqnarray}
where the amplitudes of the coherent states $\alpha_k$ and $\alpha_{-k}$ depend on the 
atomic populations of the wells and on their separation. It is immediately apparent from 
Eq.~(\ref{psitfock}) that the total optical field is not entangled with the atoms. 
As previously remarked, in the case of coherent states, which are eigenstates of the 
annihilation operator, quantum jumps do not affect the light field, hence $\ket{\psi(t)}$ 
is always a product state for this four-partite system, that is, no entanglement builds 
up between the atoms and the field.

\begin{figure}
\includegraphics{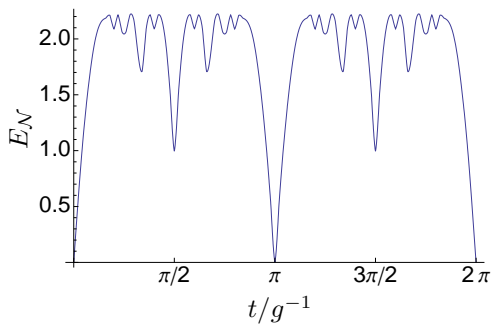}
\caption{The logarithmic negativity of light and atoms for a perfect cavity(no decay and no pump).
The initial coherent amplitude of $k$ mode $\alpha=1$ and atoms are in
superfluid states $|\psi_{\rm SF2}\rangle$ with total number of atoms $N=8$
and well separation $d=\lambda/4$. }
\label{logsf21}
\end{figure}

The situation is markedly different for atoms in a superfluid state.
For example, in the atom number-conserving approximation $\ket{\psi_{\rm SF2}}$, and in 
the absence of quantum jumps between times 0 and $t$, the state of the system at time $t$ is
\begin{eqnarray}
\ket{\psi(t)}=  \sum_{n_0}^N  b_{n_0} \ket{\alpha_k(n_0, d, t), \alpha_{-k}(n_0, d, t); n_0,N-n_0} ,\nonumber \\
\label{psitsf2}
\end{eqnarray}
a state for which the light field and atoms are clearly entangled. Dissipation and the 
associated quantum jumps will also clearly impact the amount of entanglement in that case. 
To illustrate that point, we first evaluate the logarithmic negativity of the bipartite 
atoms-optical field system for a well separation of $d=\lambda/4$, while neglecting pump 
and decay mechanisms. In that case, from Eqs.~(\ref{nd}) and (\ref{effectiveh1}) the 
system is described by the Hamiltonian
\begin{eqnarray}
\hat H&=& \hbar g[(\hat n_0+\hat n_1)(\hat a_k^\dagger \hat a_k+\hat a_{-k}^\dagger \hat a_{-k}) \nonumber \\
 &+& (\hat n_0-\hat n_1) (\hat a_{-k}^\dagger\hat a_k +\hat a_k^\dagger \hat a_{-k})],
\label{hperfect}
\end{eqnarray}
and we find easily that
\begin{eqnarray*}
\ket{\psi(t)} =\sum_{n_0}^8 \beta_{n_0} \ket{\alpha_k(n_0, t), \alpha_{-k}(n_0, t); n_0,N-n_0}
\end{eqnarray*}
where
\begin{eqnarray}
\alpha_k (n_0,n_1,t) &=& \frac{\alpha}{2} \left( e^{-2ig n_0 t}+ e^{-2ig n_1 t} \right),  \nonumber \\
\alpha_{-k} (n_0,n_1,t) &=& \frac{\alpha}{2} \left( e^{-2ig n_0t}- e^{-2ig n_1 t} \right).
\label{amp}
\end{eqnarray}
The logarithmic negativity $E_{\cal N}$ corresponding to that case is shown in 
Fig.~\ref{logsf21} for $N=8$ atoms. We observe that $E_{\cal N} = 0 $ for $t=\ell \pi/g$ 
where $\ell$ is a integer, indicating the absence of entanglement between the light and 
the atoms. These are the times when the atoms do not reflect any light into the $-k$-mode, 
and the state of the system has undergone a full revival to its original, unentangled form.

\begin{figure}
\includegraphics{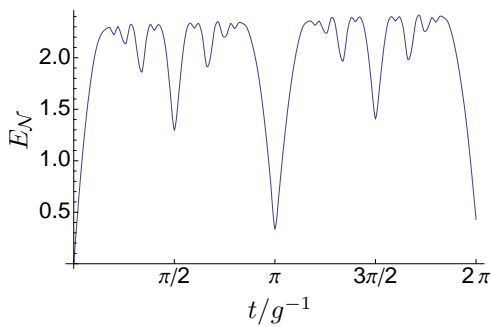}
\caption{Logarithmic negativity for a perfect cavity, without decay
but including the pump.The initial coherent amplitude of the $k$ mode is $\alpha=1$ and the atoms are initially in
the superfluid state $|\psi_{\rm SF2}\rangle$ with total number of atoms $N=8$
and well separation $d=\lambda/4$. The pump constant is $\eta=0.5 g$. }
\label{logsf22}
\end{figure}

\begin{figure}[h]
\includegraphics{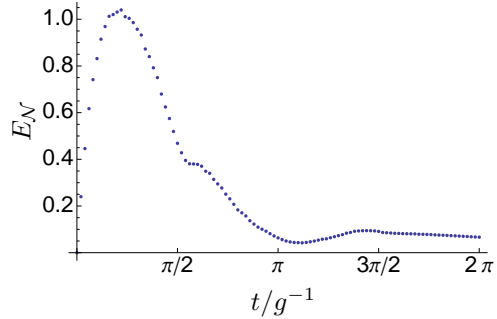}
\caption{Logarithmic negativity including dissipation.
The initial coherent amplitude of the $k$ mode is $\alpha=1$ and the atoms are in a
superfluid state $|\psi_{\rm SF2}\rangle$ with total number of atoms $N=2$
and well separation $d=\lambda/4$.
The pump constant is $\eta=0.2g$ and the decay constant is $\gamma=0.7g$.
The curve has been averaged over 100 trajectories. }
\label{logsf23}
\end{figure}

The situation is changed when including the external pump in the description of the system, 
in which case there is no longer an exact revival. Figure~\ref{logsf22} illustrates that 
a remnant of that feature is still observable in that case, but is disappears almost 
completely when both pump and dissipation are included, see Fig.~\ref{logsf23}. In that 
latter case, the entanglement between the light field and the atoms disappears completely 
over time as would be expected.

As a final point, we remark that similar results hold in case the superfluid atomic system 
is described by the state $\ket{\psi_{\rm SF1}}$, although the maximum value of the 
logarithmic negativity is now larger, due to the larger number of terms in the expansion 
of that state in terms of number states.

\section{Conclusions}

We have studied the interaction of a light field and ultracold atoms
either in the Mott insulator state or in the superfluid state in a
two-well optical lattice in a lossy cavity. We have calculated the
reflected intensity, the physical spectrum and the logarithmic
negativity, and shown that all three observables present completely different
features for the two different manybody atomic states, allowing one
can distinguish these states.

Future work will generalize this analysis to more complex manybody atomic states, using 
in particular Laguerre-Gaussian modes of the light field to characterize the state of 
vortex lattices. We are also extending this work to optomechanical situations, 
using radiation pressure force on Fabry-P{\'e}rot resonators with moving mirrors to 
induce quantum phase transitions and monitor them optically in real time.

\section*{Acknowledgements}
We thank D. Meiser and M. Bhattacharya for stimulating discussions.
This work is supported in part by the US Office of Naval Research,
by the National Science Foundation, and by the US Army Research
Office.

\bibliography{mybibliography}
\end{document}